\documentclass[aps,prd,twocolumn,showpacs,nofootinbib]{revtex4-1}

\usepackage{graphicx}
\usepackage{amsmath}
\usepackage{amssymb}
\usepackage{bm}
\usepackage{bbm}
\usepackage{color}
\usepackage{slashed}
\usepackage[colorlinks=true,linkcolor=blue,citecolor=blue, urlcolor=blue]{hyperref}

\begin{document}

\title{The NLO fragmentation functions of heavy quarks into heavy quarkonia}

\author{Xu-Chang Zheng$^{a}$}
\email{zhengxc@cqu.edu.cn}
\author{Chao-Hsi Chang$^{b,c,d}$}
\email{zhangzx@itp.ac.cn}
\author{Xing-Gang Wu$^{a}$}
\email{wuxg@cqu.edu.cn}

\affiliation{$^a$ Department of Physics, Chongqing University, Chongqing 401331, P.R. China.\\
$^b$ Key Laboratory of Theoretical Physics, Institute of Theoretical Physics, Chinese Academy of Sciences, Beijing 100190, China.\\
$^c$ School of Physical Sciences, University of Chinese Academy of Sciences, Beijing 100049, China.\\
$^d$ CCAST (World Laboratory), Beijing 100190, China.}

\begin{abstract}

In the paper, we derive the next-to-leading order (NLO) fragmentation function for a heavy quark, either charm or bottom, into a heavy quarkonium $J/\Psi$ or $\Upsilon$. The ultra-violet divergences in the real corrections are removed through the operator renormalization, which is performed under the modified minimal subtraction scheme. We then obtain the NLO fragmentation function at an initial factorization scale, e.g. $\mu_{F}=3 m_c$ for $c\to J/\Psi$ and $\mu_{F}=3m_b$ for $b\to \Upsilon$, which can be evolved to any scale via the use of Dokshitzer-Gribov-Lipatov-Altarelli-Parisi equation. As an initial application of those fragmentation functions, we study the $J/\Psi$ ($\Upsilon$) production at a high luminosity $e^+e^-$ collider which runs at the energy around the $Z$ pole and could be a suitable platform for testing the fragmentation function.

\end{abstract}


\maketitle

\section{Introduction}

Since the observation of $J/\Psi$ meson, the heavy quarkonium has attracted great interests from theorists and experimentalists. Due to the constituent quark ($Q$) and antiquark ($\bar{Q}$) of a heavy quarkonium are heavy, i.e, $m_Q \gg \Lambda_{QCD}$, its production rate and decay width involve both perturbative and non-perturbative aspects of quantum chromodynamics (QCD). It then provides a good platform for testing various QCD factorization theories. Within the framework of the nonrelativistic QCD (NRQCD) factorization theory~\cite{nrqcd}, the production cross section of a heavy quarkonium via the collision of two incident particles $A$ and $B$ can be written as
\begin{eqnarray}
&&d \sigma(A+B\to H+X)\nonumber \\
&&=\sum_n d \tilde{\sigma}(A+B \to (Q\bar{Q})[n]+X) \langle {\cal O}_n^H \rangle,
\label{nrqcdfact}
\end{eqnarray}
where $H$ denotes the produced heavy quarkonium, $d \tilde{\sigma}$ is the production cross section for the perturbative state $(Q\bar{Q})[n]$ with quantum numbers $n$, which is calculable and can be expanded in powers of the strong coupling constant $\alpha_s$. $\langle {\cal O}_n^H \rangle$ denotes the non-perturbative but universal NRQCD matrix element, which is proportional to the transition probability of the perturbative state $(Q\bar{Q})[n]$ into the hadron state $H$.

The NRQCD factorization formulism has been used to deal with the quarkonium production at $e^+e^-$ and hadronic colliders, and most of the calculations have been performed up to next-to-leading order (NLO) accuracy~\cite{ybook1, ybook2}. In some cases, there are large logarithms in the short-distance part $d \tilde{\sigma}$. For instance, there are large logarithms in powers of $\ln (\sqrt{s}/m_Q)$ for the heavy quarkonium production at a $e^+ e^-$ collider with $\sqrt{s}$ being the central of mass collision energy, or there are large logarithms in powers of $\ln (p_T/m_Q)$ in high $p_T$ region ($p_T$ being the  transverse momentum of the quarkonium) at a hadronic collider. Those large logarithms may spoil the convergence of the perturbative expansion, leading to the unreliable pQCD predictions. As a solution, it has been pointed out that those logarithms mainly come from the emission of collinear gluons, which can be systematically resummed through the Dokshitzer-Gribov-Lipatov-Altarelli-Parisi (DGLAP) evolution equation~\cite{dglap}.

In the present paper, we shall take the heavy quarkonium production at an $e^+e^-$ collider as an explicit example to explain this idea, especially, we shall give the results for the heavy quark to heavy quarkonium fragmentation function up to NLO accuracy. The fragmentation function that gives the probability for the
splitting of a parton into the desired hadron plus other partons. For a hadron composed of only heavy quarks, its fragmentation function can be calculated using pQCD theory. The Z boson decays $Z\to J/\Psi+X$ and $Z\to \Upsilon+X$ have been studied up to NLO level using the complete
pQCD approach \cite{JXWang}, and in the present paper we will study these processes but using the fragmentation approach. Compared with the complete pQCD approach, even
though the fragmentation approach neglects some power-suppressed terms, it resums the large collinear logarithms
more conveniently via the use of the DGLAP evolution
equation and achieves more reliable predictions in specific
kinematic regions.

We shall also study these processes but using the fragmentation function approach. Compared with the complete pQCD approach, the fragmentation function approach can resum the large collinear logarithms, however, it neglects some power suppressed terms. One may combine these two approaches to give a good prediction.

Under the pQCD factorization theory, the production cross section of heavy quarkonium at an $e^+ e^-$ collider can be factorized as
\begin{eqnarray}
 d \sigma(e^+ e^- \to H(p)+X)=&& \sum_i d \hat{\sigma}(e^+ e^- \to i(p/z)+X,\mu_F) \nonumber \\
&&  \otimes D_{i\to H}(z,\mu_F)+{\cal O}(m_Q^2/s),
\label{pqcdfact}
\end{eqnarray}
where the sum extends over all the parton types, $d \hat{\sigma}$ denotes the partonic cross section (coefficient function), $D_{i\to H}(z,\mu_F)$ denotes the fragmentation function (decay function) for the parton $i$ into heavy quarkonium $H$ with longitudinal momentum fraction $z$. $\mu_F$ is the factorization scale which separates the energy scales of two parts. In order to avoid large logarithms appearing in $d \hat{\sigma}$, $\mu_F$ is usually set as $\mu_F={\cal O}(\sqrt{s})$.

The pQCD factorization formula (\ref{pqcdfact}) was firstly suggested by Collins and Soper for the inclusive production of a light hadron~\cite{Collins}, and the proof of the pQCD factorization formula to the case of quarkonium production was given by Nayak, Qiu, and Sterman~\cite{fragnrqcd}. The recent progress of the pQCD factorization for quarkonium production is the derivation of the next-leading-power (NLP) contribution, which comes from the double-parton fragmentation~\cite{nlp1, nlp2, nlp3, nlp4}. The fragmentation function $D_{i\to H}(z,\mu_F)$ contains nonperturbative information, which is calculable through the NRQCD factorization (or the Mandelstam formulation~\cite{Mandelstam} under the instantaneous approximation)~\cite{fragbc1, fragbc2}, e.g. $D_{i\to H}(z,\mu_F)$ can be factorized as
\begin{eqnarray}
D_{i\to H}(z,\mu_F)=\sum_n d_{i\to (Q\bar{Q})[n]}(z,\mu_F) \langle {\cal O}_n^H \rangle,
\label{frag.nrqcd}
\end{eqnarray}
where $d_{i\to (Q\bar{Q})[n]}(z,\mu_F)$ is the short-distance coefficient, which contains the logarithms of $\mu_F/m_Q$. To avoid such kind of large logarithms, one can first calculate the fragmentation function at some initial factorization scale which is of order ${\cal O} (m_Q)$, and then evolve it to a higher factorization scale by using the Dokshitzer-Gribov-Lipatov-Altarelli-Parisi (DGLAP) evolution equation~\cite{dglap1, dglap2, dglap3}:
\begin{eqnarray}
&&\frac{d}{d~{\rm ln} \mu_F^2}D_{i \to H}(z,\mu_F)\nonumber \\
&&=\frac{\alpha_s(\mu_F)}{2\pi}\sum_j \int_z^1 \frac{dy}{y}P_{ji}(y,\alpha_s(\mu_F)) D_{j \to H}(z/y,\mu_F),
\label{dglap}
\end{eqnarray}
where $P_{ji}$ are splitting functions, which can be expanded in perturbative series as:
\begin{eqnarray}
P_{ji}(y,\alpha_s(\mu_F))=&& \frac{\alpha_s(\mu_F)}{2\pi}P^{(0)}_{ji}(y)+\frac{\alpha^2_s(\mu_F)}{(2\pi)^2} P^{(1)}_{ji}(y)\nonumber \\
&&+{\cal O}(\alpha_s^3).
\label{spfun}
\end{eqnarray}
For the quark to quark case, the LO coefficient
\begin{equation}
P^{(0)}_{QQ}(y)=C_F\left[\frac{1+y^2}{(1-y)_+}+\frac{3}{2}\delta(1-y)\right],
\label{lospfun}
\end{equation}
where $C_F=4/3$ for $SU(3)_c$ group. The NLO coefficient $P^{(1)}_{QQ}(y)$ is too lengthy to be presented here, which can be found in Refs.\cite{nlospfun1, nlospfun2, nlospfun3}.

The LO fragmentation function for a heavy quark to $J/\Psi$ or $\Upsilon$ was firstly calculated by Braaten and Cheung in 1993~\cite{Braaten1}, where the LO fragmentation function is derived from a LO calculation of the process $Z \to J/\Psi+c+\bar{c}$. Subsequently, Ma calculated the LO fragmentation functions~\cite{Ma} by using the gauge invariant definition suggested by Collins and Soper. In Ref.\cite{Sepahvand}, the authors calculated the NLO corrections to the transverse-momentum dependent fragmentation functions for a heavy quark to $J/\Psi$ and $\Upsilon$, which is however not convenient for practical applications. Recently, the NLO fragmentation function for a gluon into heavy quarkonium has been finished by Refs.\cite{Braaten2, Braaten3, YJia, YQMa}. In the present paper, we shall give the fragmentation functions for a heavy quark into $J/\Psi$ and $\Upsilon$ up to NLO level.

The remaining parts of the paper are organized as follows. In Sec.II, we present the LO fragmentation function for a heavy quark into the heavy quarkonium $J/\Psi$ or $\Upsilon$. In Sec.III, we present the NLO fragmentation functions $D_{c \to J/\Psi}$ and $D_{b \to \Upsilon}$, in which the renormalization is carried out by using the conventional $\overline{\rm MS}$-scheme. In Sec.IV, we apply those NLO fragmentation functions to $J/\Psi$ and $\Upsilon$ production at a super $Z$ factory. Sec.V is reserved for a summary.

\section{The LO fragmentation function}
\label{LOFrag}

Before carrying out the calculation for the fragmentation function, we first give a brief introduction of the gauge invariant fragmentation function suggested by Collins and Soper~\cite{Collins}. We adopt the dimensional regularization to regularize the infrared (IR) and ultraviolet (UV) divergences, and shall work in $d = 4-2\epsilon$ dimensional space-time.

The light-cone coordinates are conventionally adopted to define the fragmentation function, where a vector $V^{\mu}$ in $d$ dimensions is expressed as $V^{\mu}=(V^+,V^-,\textbf{V}_{\perp})=((V^0+V^{d-1})/\sqrt{2},(V^0-V^{d-1})/\sqrt{2},\textbf{V}_{\perp})$. The scalar
product of two vectors $V$ and $W$ then becomes $V\cdot W=V^+ W^-+V^- W^+-\textbf{V}_{\perp} \cdot \textbf{W}_{\perp}$. The gauge-invariant fragmentation function for the heavy quark $Q$ into a spin-triplet and color-singlet quarkonium $H$ in $d=4-2\epsilon$ dimension is defined as
\begin{widetext}
\begin{eqnarray}
D_{Q\to H}(z)=&&\frac{z^{d-3}}{2\pi}\sum_{X} \int dx^- e^{-iP^+ x^-/z} \nonumber \\
&&\times \frac{1}{N_c} {\rm Tr}_{\rm color}  \frac{1}{4} {\rm Tr}_{\rm Dirac} \left\lbrace \gamma^+ \langle 0 \vert \Psi(0)\bar{{\cal P}} {\rm exp}\left[ig_s \int_{0}^{\infty} dy^- A_a^+(0^+,y^-,0_\perp)t_a^T \right]\vert H(P^+,0_\perp)+X \rangle \right. \nonumber \\
&&\left. \times\langle H(P^+,0_\perp)+X\vert {\cal P} {\rm exp}\left[-ig_s \int_{x^-}^{\infty} dy^- A_a^+(0^+,y^-,0_\perp)t_a^T \right] \bar{\Psi}(x)\vert 0\rangle\right\rbrace,
\label{defrag1}
\end{eqnarray}
\end{widetext}
where $\Psi$ and $A_a^{\mu}$ are quark field and gluon field, respectively. $t^a$ is the color matrix, ${\cal P}$ implies the path ordering, $z \equiv P^+/K^+$ is the longitudinal momentum fraction carried from the incident heavy quark $Q$. This definition is carried out in a reference frame where the quarkonium $H$ carries no transverse momentum, i.e, $P^{\mu}=(P^+,P^-=m_H^2/2P^+,0_\perp)$. It is convenient to introduce a light-like vector $n^\mu$, which has the value of $n^\mu=(0,1,0_\perp)$ in the reference frame where the definition of the fragmentation function is carried out. In this frame, the plus component of a momentum $p$ can be expressed as $p^+=p \cdot n$, and $z=P\cdot n/K \cdot n$. The Feynman rules for the fragmentation function can be directly derived from the above definition~\cite{Bcfragnlo}.

The fragmentation function (\ref{defrag1}) is gauge independent. For practical treatment, we work under the usual Feynman gauge. To derive the fragmentation function $D_{Q\to H}$, we first calculate the fragmentation function for the production of a free on-shell $Q\bar{Q}$ state with the quantum numbers $^3S^{[1]}_1$ (i.e, $D_{Q\to Q\bar{Q}[^3S_1^{[1]}]}$), where the superscript ``[1]" denotes the $Q\bar{Q}$ state is in color singlet. Then, $D_{Q\to H}$ is obtained by replacing the NRQCD matrix element of a free $Q\bar{Q}$ state ($\langle {\cal O} ^{Q\bar{Q}[^3S_1^{[1]}]}(^3S_1^{[1]})\rangle$) to the matrix element of the quarkonium ($\langle {\cal O} ^{H}(^3S_1^{[1]})\rangle$).

\begin{figure}[htbp]
\includegraphics[width=0.5\textwidth]{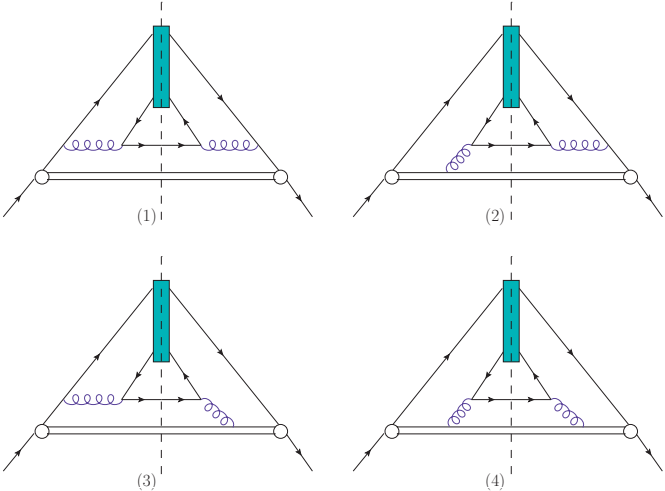}
\caption{The LO cut diagrams for the fragmentation function $D_{Q\to Q\bar{Q}[^3S_1^{[1]}]}$. The double line stands for the Wilson line which ensures the gauge invariance of the squared amplitude.
 } \label{feylo}
\end{figure}

At the LO level, we need to deal with process $Q(K)\to Q\bar{Q}[^3S_1^{[1]}](p_1)+Q(p_2)$. As shown by Fig.\ref{feylo}, there are four LO cut diagrams which contribute to $D_{Q\to Q\bar{Q}[^3S_1^{[1]}]}$. The four squared amplitudes from four cut diagrams are
\begin{eqnarray}
{\cal A}_1=&& {\rm tr}\left[(\slashed{p}_{2}+m_Q)(ig_s\gamma^{\mu}t^a)\Pi \Lambda_1 (ig_s\gamma_{\mu}t^a)  \right. \nonumber \\
&& \cdot   \left. \frac{i}{\slashed{p_1}+\slashed{p_2}-m_Q+i\epsilon} \slashed{n} \frac{-i}{\slashed{p_1}+\slashed{p_2}-m_Q-i\epsilon}   \right. \nonumber \\
&& \cdot   (-ig_s\gamma^{\nu}t^b)\bar{\Pi}\Lambda_1 (-ig_s\gamma_{\nu}t^b) \Big]\nonumber \\
&& \cdot \frac{-i}{(p_{1}/2+p_2)^2+i\epsilon} \frac{i}{(p_{1}/2+p_2)^2-i\epsilon} ,
\end{eqnarray}
\begin{eqnarray}
{\cal A}_2=&& {\rm tr}\left[(\slashed{p}_{2}+m_Q)(ig_s\gamma^{\mu}t^a)\Pi \Lambda_1   \right. \nonumber \\
&& \cdot \frac{i}{(-p_1/2-p_2)\cdot n+i\epsilon} (ig_s n_{\mu}t^a)\slashed{n}\nonumber \\
&& \left. \cdot \frac{-i}{\slashed{p_1}+\slashed{p_2}-m_Q-i\epsilon}  (-ig_s\gamma^{\nu}t^b)\bar{\Pi}\Lambda_1 (-ig_s\gamma_{\nu}t^b) \right]\nonumber \\
&&\cdot  \frac{-i}{(p_{1}/2+p_2)^2+i\epsilon} \frac{i}{(p_{1}/2+p_2)^2-i\epsilon},
\end{eqnarray}
\begin{eqnarray}
{\cal A}_3=&& {\rm tr}\left[(\slashed{p}_{2}+m_Q)(ig_s\gamma^{\mu}t^a)\Pi \Lambda_1 (ig_s\gamma_{\mu}t^a)
 \right. \nonumber \\
&& \cdot  \frac{i}{\slashed{p_1}+\slashed{p_2}-m_Q+i\epsilon} \slashed{n} (-ig_s n_{\nu}t^b)  \nonumber \\
&& \left. \cdot  \frac{-i}{(-p_1/2-p_2)\cdot n-i\epsilon}\bar{\Pi}\Lambda_1 (-ig_s\gamma_{\nu}t^b) \right] \nonumber \\
&& \cdot \frac{-i}{(p_{1}/2+p_2)^2+i\epsilon} \frac{i}{(p_{1}/2+p_2)^2-i\epsilon} ,\\
{\cal A}_4=&& {\rm tr}\left[(\slashed{p}_{2}+m_Q)(ig_s\gamma^{\mu}t^a)\Pi \Lambda_1   \right. \nonumber \\
&& \cdot \frac{i}{(-p_1/2-p_2)\cdot n+i\epsilon} (ig_s n_{\mu}t^a)\slashed{n}
 (-ig_s n_{\nu}t^b)  \nonumber \\
&& \cdot \left.  \frac{-i}{(-p_1/2-p_2)\cdot n-i\epsilon} \bar{\Pi}\Lambda_1 (-ig_s\gamma_{\nu}t^b) \right]\nonumber \\
&& \cdot \frac{-i}{(p_{1}/2+p_2)^2+i\epsilon} \frac{i}{(p_{1}/2+p_2)^2-i\epsilon} .
\end{eqnarray}
Here $\Pi$ denotes the spin projector, and for $^3S_1$ state
\begin{eqnarray}
\Pi= -\frac{1}{{2\sqrt{2 m_Q}}}\slashed{\epsilon}(p_1) (\slashed{p}_{1} + 2 m_Q),
\end{eqnarray}
and $\bar{\Pi}\equiv \gamma^0 \Pi^{\dagger} \gamma^0$. $\Lambda_1$ is color-singlet prejector
\begin{eqnarray}
\Lambda_1= \frac{\textbf{1}}{\sqrt{3}},
\end{eqnarray}
where $\textbf{1}$ is the ${\rm SU}_{c}(3)$ unit matrix.

Then we obtain the total squared amplitude at the LO level,
\begin{eqnarray}
{\cal A}_{\rm Born}&=& \sum_{j=1}^{4}{\cal A}_j = \frac{2C_F^2 g_s^4  K\cdot n}{ (2-z)^2 m_Q}\sum_{j=2}^4 \frac{a_j m_Q^{2(j-2)}}{(s_1-m_Q^2)^j},\label{aborn}
\end{eqnarray}
where $s_1=(p_1+p_2)^2$ is the invariant mass of final $Q\bar{Q}[^3S^{[1]}_1]+Q$, and the coefficients $a_{j=2,3,4}$ are
\begin{eqnarray}
a_2=&&(1-z)[(d^3-17 d^2+100 d-156)z^2-4(d^3-13 d^2\nonumber \\
&&+56 d-84)z+4(d^3-9 d^2+28 d-28)], \nonumber \\
a_3=&&8(z-2)[(d^2-9 d+16) z^2-2 (d^2-5 d+16) z+16], \nonumber \\
a_4=&&-64 (d-1) (z-2)^2. \nonumber
\end{eqnarray}

The differential phase space for the LO fragmentation function is
\begin{eqnarray}
d\phi_{\rm Born} = \frac{dp_2^+}{2p_2^+}\frac{ d^{d-2}\textbf{p}_{2\perp}}{(2\pi)^{d-2}}2\pi \delta(K^+-p_1^+-p_2^+),  \label{phase}
\end{eqnarray}
where the $\delta$-function comes from the final cut of the eikonal line. The integration over $p_2^+$ can be carried out with the $\delta$-function. The integration over the angles of $\textbf{p}_{2\perp}$ is trivial and can be carried out easily. Then we have
\begin{eqnarray}
d\phi_{\rm Born}=&& \frac{z^{-1+\epsilon}(1-z)^{-\epsilon}}{2(4\pi)^{1-\epsilon}\Gamma(1-\epsilon)K\cdot n} \nonumber \\
&&\times \left(s_1-\frac{4m_Q^2}{z}-\frac{m_Q^2}{1-z}\right)^{-\epsilon}ds_1.\label{phslo}
\end{eqnarray}
The range of $s_1$ is from $[4m_Q^2/z+m_Q^2/(1-z)]$ to $+\infty$.

The LO fragmentation function for $Q\to Q\bar{Q}[^3S^{[1]}_1]$ can be obtained through
\begin{eqnarray}
D^{\rm LO}_{Q\to Q\bar{Q}[^3S_1^{[1]}]}(z)=N_{CS}\int d\phi_{\rm Born} {\cal A}_{\rm Born},
\end{eqnarray}
where $N_{CS}=z^{1-2\epsilon}/8\pi N_c$ is an overall factor. Performing the integration over $s_1$, we obtain
\begin{eqnarray}
&&D^{\rm LO}_{Q\to Q\bar{Q}[^3S_1^{[1]}]}(z) \nonumber \\
&&=\frac{C_F^2\alpha_s^2 z(1-z)(4\pi)^{\epsilon}\Gamma(1+\epsilon)}{2N_c (2-z)^{4+2\epsilon} m_Q^{3+2\epsilon}}\left[a_2+a_3\frac{(1+\epsilon)z(1-z)}{2(2-z)^2} \right. \nonumber \\
&&~~\left. +a_4\frac{(2+\epsilon)(1+\epsilon)z^2(1-z)^2}{6(2-z)^4}\right].
\end{eqnarray}
Setting $d=4$, we obtain
\begin{eqnarray}
&&D^{\rm LO}_{Q\to Q\bar{Q}[^3S_1^{[1]}]}(z)\nonumber \\
&&= \frac{32\alpha_s^2 z(1-z)^2}{27(2-z)^6 m_Q^3}(5 z^4-32 z^3+72 z^2-32 z+16) \nonumber \\
&&~~\times \frac{\langle {\cal O} ^{Q\bar{Q}[^3S_1^{[1]}]}(^3S_1^{[1]})\rangle}{6N_c}\label{lobc*}.
\end{eqnarray}
Here, the LO fragmentation function for the free $Q\bar{Q}[^3S_1^{[1]}]$ state has been written as the factorization form by the use of the fact that
\begin{eqnarray}
\langle {\cal O} ^{Q\bar{Q}[^3S_1^{[1]}]}(^3S_1^{[1]})\rangle=2(d-1)N_c.
\end{eqnarray}
at the order $\alpha_s^0$ with the normalization of the NRQCD matrix element in Ref.\cite{nrqcd}. Then the LO fragmentation function for the quarkonium can be obtained through replacing $\langle {\cal O} ^{Q\bar{Q}[^3S_1^{[1]}]}(^3S_1^{[1]})\rangle$ by $\langle {\cal O} ^{H}(^3S_1^{[1]})\rangle$. Under the leading non-relativistic approximation, the matrix element $\langle {\cal O} ^{H}(^3S_1^{[1]})\rangle$ can be expressed by the radial wave function at the origin for the quarkonium $H$, i.e,
\begin{eqnarray}
\langle {\cal O} ^{H}(^3S_1^{[1]})\rangle \approx (d-1)N_c\vert R_S(0)\vert^2/(2\pi). \label{ldme}
\end{eqnarray}
Finally, the LO fragmentation function for the $^3S_1$ quarkonium state takes the form
\begin{eqnarray}
&&D^{\rm LO}_{Q\to H}(z)\nonumber \\
&&= \frac{8\alpha_s^2 z(1-z)^2\vert R_S(0)\vert^2}{27\pi (2-z)^6 m_Q^3}(5 z^4-32 z^3+72 z^2-32 z+16), \nonumber \\
\label{lofrag}
\end{eqnarray}
which is exactly the same as that of Ref.\cite{Braaten1}.

\section{The NLO correction to the fragmentation function}
\label{sec3}

At the NLO level, we need to deal with the virtual and real corrections to the LO terms. It is hard to give the analytic expressions for those NLO-terms. In the following subsections, we shall give some explanations on how to deal with the virtual and real corrections.

In doing the calculations, the FeynCalc package~\cite{feyncalc1, feyncalc2} is adopted to carry out the color and Dirac traces, the \$Apart package~\cite{apart} and the FIRE package~\cite{fire} are used to do partial fraction and integration-by-parts (IBP) reduction. The master integrals are calculated by using the LoopTools package~\cite{looptools}. As a subtle point, there are some master integrals, which contain an eikonal propagator and can not be calculated by using the LoopTools, and we adopt the method introduced in Ref.\cite{Braaten2} to deal with those master integrals.

\subsection{The virtual corrections}

\begin{figure}[htbp]
\includegraphics[width=0.5\textwidth]{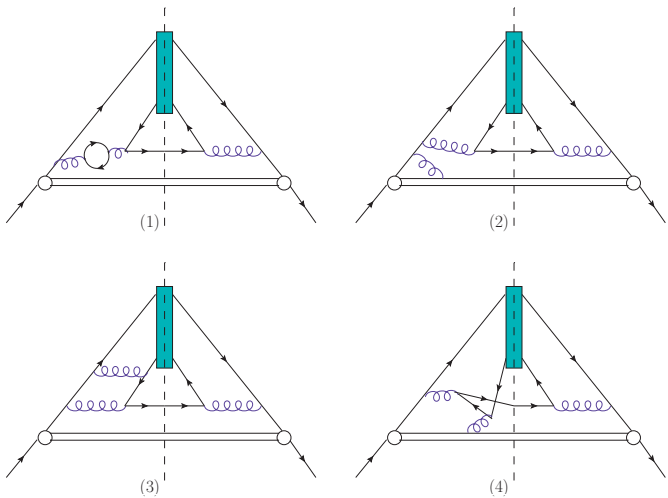}
\caption{Four typical virtual corrections to the fragmentation function $D_{Q\to Q\bar{Q}[^3S_1^{[1]}]}$. The double line stands for the Wilson line which ensures the gauge invariance of the squared amplitude. } \label{feyvir}
\end{figure}

At the NLO level, the virtual corrections come from the cut diagrams containing a loop on either side of the cut. Four typical virtual corrections are shown in Fig.\ref{feyvir}. The differential phase space of the virtual corrections is the same as that of the LO one, e.g. Eq.(\ref{phase}). The virtual corrections can be obtained through
\begin{eqnarray}
D^{\rm virtual}_{Q\to Q\bar{Q}[^3S_1^{[1]}]}(z)=N_{CS}\int d\phi_{\rm Born} {\cal A}_{\rm virtual}.
\end{eqnarray}
where ${\cal A}_{\rm virtual}$ is the squared amplitude for the virtual corrections.

We adopt the method of regions~\cite{region} to calculate the fragmentation function. Within this method, we only need to calculate the contributions from the hard region, the Coulomb divergences which come from the potential region do not appear in the calculations. The hard region contributions are then obtained by expanding the relative velocity between the produced $Q$ and $\bar{Q}$ before the loop integration. Thus in the calculation, we only need to set the relative momentum to zero before the loop integration~\cite{region}.

\subsection{The real corrections}

\begin{figure}[htbp]
\includegraphics[width=0.5\textwidth]{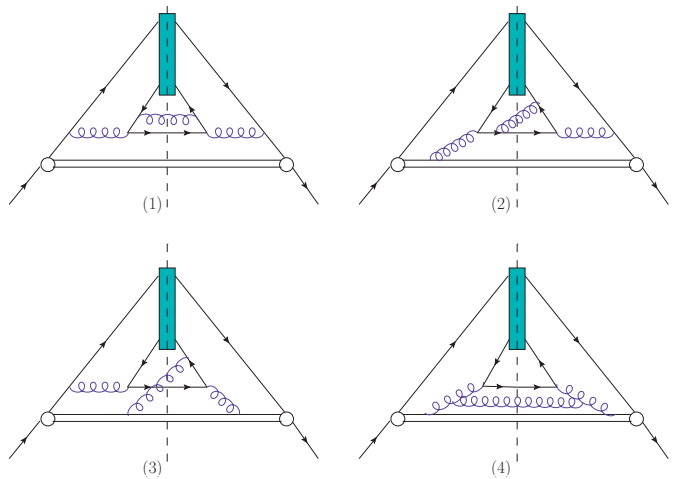}
\caption{Four typical real corrections to the fragmentation function $D_{Q\to Q\bar{Q}[^3S_1^{[1]}]}$. The double line stands for the Wilson line which ensures the gauge invariance of the squared amplitude. } \label{feyreal}
\end{figure}

At the NLO level, the real corrections come from the fragmentation process emitting an extra gluon, i.e, we need to deal with the process, $Q(K)\to Q\bar{Q}[^3S_1^{[1]}](p_1)+Q(p_2)+g(p_3) $. The cut diagrams for the real corrections can be obtained from the LO cut diagrams by adding a gluon line crossing the cut. Four typical real correction cut diagrams are shown in Fig.\ref{feyreal}.

The differential phase space for the real correction is
\begin{eqnarray}
d\phi_{\rm real}=&&2\pi \delta\left(K^+ - \sum_{i=1}^{3}  p_i^+\right)  \prod_{i=2,3}\frac{dp_i^+}{2p_i^+}\frac{ d^{d-2}\textbf{p}_{i\perp}}{(2\pi)^{d-2}}.
\end{eqnarray}
The real corrections can be obtained through
\begin{eqnarray}
D^{\rm real}_{Q\to Q\bar{Q}[^3S_1^{[1]}]}(z)=N_{CS}\int d\phi_{\rm real} {\cal A}_{\rm real},
\end{eqnarray}
where ${\cal A}_{\rm real}$ denotes the squared amplitude for the real corrections. As for the real corrections, the IR divergences come from the limits $p_3 \to 0$ and $p_3 \cdot n \to 0$, and the UV divergences come from the limit $\vert \textbf{p}_{3\perp} \vert \to \infty$. We adopt the method of Ref.\cite{Braaten2} to extract those divergences. Following this method, the real corrections can be represented as
\begin{eqnarray}
D^{\rm real}_{Q \to Q\bar{Q}[n]}(z)=&&N_{CS}\int d\phi_{\rm real} ({\cal A}_{\rm real}-{\cal A}_S)\nonumber \\
&&+N_{CS}\int d\phi_{\rm real} {\cal A}_S.\label{dsub}
\end{eqnarray}
where ${\cal A}_S$ denotes the subtraction term, which has the same singularities as the squared amplitude of the real corrections. The first term on the right hand side of Eq.(\ref{dsub}) is finite and can be directly calculated in $4$-dimensions. The integral of the subtraction term is divergent and should be calculated in $d$-dimensions.

The subtraction term can be constructed according to the singularity behavior of the squared amplitude of the real corrections. More explicitly, the squared amplitude for the real corrections can be written as
\begin{widetext}
\begin{eqnarray}
{\cal A}_{\rm real}=&&\frac{b_1(s_1,z)}{(1-y)(s-m_Q^2)}+\frac{b_2(s_1,z)}{(1-y)(s_2-m_Q^2)}+\frac{b_3(s_1,z)}{(1-y)s_3}+\frac{c_1(s_1,z,y)}{s-m_Q^2}+\frac{c_2(s_1,z,y)p_1\cdot p_3}{(s-m_Q^2)^2}+\frac{c_3(s_1,z,y)}{s_2-m_Q^2}\nonumber \\
&&+\frac{c_4(s_1,z,y)p_2\cdot p_3}{(s_2-m_Q^2)^2} +\frac{c_5(s_1,z,y)}{s_3}+\frac{c_6(s_1,z,y)p_1\cdot p_3}{s_3^2}+\frac{d_1(s_1,z)(1-u)(s_1-m_Q^2)}{u~t_1(s-m_Q^2)}\nonumber \\
&&+\frac{d_2(s_1,z)(1-u)(s_1-m_Q^2)}{2~u~t_1~s_3}+\frac{d_3(s_1,z)(1-u)(s_1-m_Q^2)^2}{2~u~t_1~s_3(s-m_Q^2)}+\frac{d_4(s_1,z)(s_1-m_Q^2)^2}{2~u~t_2 (s-m_Q^2)s_3}\nonumber \\
&&+\frac{d_5(s_1,z)(s_1-m_Q^2)}{u~t_2 (s-m_Q^2)}+\frac{d_6(s_1,z)(s_1-m_Q^2)}{2~u~t_2~s_3}+\frac{g(s_1,z)(s_1-m_Q^2)^2}{2~u(s-m_Q^2)s_3}+\frac{h(s_1,z)}{t_2^2} +{\cal A}^{\rm finite}_{\rm real},
\end{eqnarray}
where the Lorentz invariants are defined as follows
\end{widetext}
\begin{eqnarray}
&&y=\frac{(p_1+p_2)\cdot n}{(p_1+p_2+p_3) \cdot n},\, u=\frac{p_3\cdot n}{(p_2+p_3) \cdot n}, \nonumber \\
&&s=(p_1+p_2+p_3)^2,~~~~ s_2=(p_{1}/2+p_3)^2, \nonumber \\
&&s_3=(p_{1}/2+p_2+p_3)^2,~~t_1=2 p_1 \cdot p_3 ,\nonumber \\
&&t_2=2 p_2 \cdot p_3.\label{defvar}
\end{eqnarray}
The ${\cal A}^{\rm finite}_{\rm real}$ stands for the terms which are finite after the phase-space integration. There are neither $1/t_1^2$ term nor $1/t_1 t_2$ term in ${\cal A}_{\rm real}$ because those terms are canceled after summing all the terms of the real corrections. The subtraction term ${\cal A}_{\rm S}$ can then be constructed as follows
\begin{widetext}
\begin{eqnarray}
{\cal A}_{\rm S}=&&\frac{b_1(s_1,z)}{(1-y)(s-m_Q^2)}+\frac{b_2(s_1,z)}{(1-y)(s_2-m_Q^2)}+\frac{b_3(s_1,z)}{(1-y)s_3}+\frac{c_1(s_1,z,y)}{s} \nonumber \\
&&+\frac{c_2(s_1,z,y)}{s^2}\left[p_1\cdot p_3-\frac{z}{2y}\left(1-\frac{2}{y}\right)s_1-\frac{1-y}{2y}(s_1+3m_Q^2)\right]+\frac{c_3(s_1,z,y)}{s_2}\nonumber \\
&&+\frac{c_4(s_1,z,y)}{s_2^2}\left[p_2\cdot p_3 +\frac{(y-z)m_Q^2}{z}\left(1+\frac{4(1-y)}{z}\right)-\frac{1-y}{2z}(s_1-5m_Q^2)\right]+\frac{c_5(s_1,z,y)}{s_3}\nonumber \\
&&+\frac{c_6(s_1,z,y)}{s_3^2}\left[p_1\cdot p_3+\frac{ z(2-z)/4-(1-y)(y-z)}{2(y-z/2)^2}(s_1-m_Q^2) \right]+\frac{d_1(\tilde{s},z)(1-u)(\tilde{s}-m_Q^2)}{u~t_1(\tilde{s}-m_Q^2+t_1/z)}\nonumber \\
&&+\frac{d_2(\tilde{s},z)(1-u)(\tilde{s}-m_Q^2)}{u~t_1[\tilde{s}-m_Q^2+(2- z)t_1/z]}+\frac{d_3(\tilde{s},z)(1-u)(\tilde{s}-m_Q^2)^2}{u~t_1(\tilde{s}-m_Q^2+t_1/z)[\tilde{s}-m_Q^2+(2-z)t_1/z]}\nonumber \\
&&+\frac{d_4(\tilde{s},z)(\tilde{s}-m_Q^2)^2}{u~t_2[\tilde{s}-m_Q^2+t_2/(1-z)][\tilde{s}-m_Q^2+(2-z)t_2/(1-z)]}+\frac{d_5(\tilde{s},z)(\tilde{s}-m_Q^2)}{u~t_2(\tilde{s}-m_Q^2+t_2/(1-z))}\nonumber \\
&&+\frac{d_6(\tilde{s},z)(\tilde{s}-m_Q^2)}{u~t_2[\tilde{s}-m_Q^2+(2-z)t_2/(1-z)]}+\frac{g(\tilde{s},z)(\tilde{s}-m_Q^2)^2} {u[\tilde{s}-m_Q^2+t_2/(1-z)][\tilde{s}-m_Q^2+(2-z)t_2/(1-z)]} +\frac{h(\tilde{s},z)}{t_2^2},
\end{eqnarray}
\end{widetext}
where $\tilde{s}$ is defined as
\begin{eqnarray}
\tilde{s}=(p_1+\tilde{p})^2
\end{eqnarray}
with
\begin{eqnarray}
\tilde{p}^{\mu}=p_2^{\mu}+p_3^{\mu}-\frac{p_2\cdot p_3}{(p_2+p_3)\cdot n} n^{\mu}.
\end{eqnarray}
Numerically, we have found that the integration of $({\cal A}_{\rm real}-{\cal A}_{\rm S})$ over the phase space is finite in $4$-dimensions, which confirms our present choice of the subtraction term ${\cal A}_{\rm S}$. Due to its much simpler structure, following the method of Ref.\cite{Bcfragnlo}, the phase-space integration over the subtraction term ${\cal A}_{\rm S}$ can be done analytically. The interesting reader can turn to Ref.\cite{Bcfragnlo} for details.

\subsection{Renormalization}

The UV divergences in the virtual and real corrections should be canceled through renormalization. The counter-term approach is adopted to carry out the renormalization, where the fragmentation function is calculated with the renormalized quark mass $m_Q$, the renormalized field $\Psi_r$, the renormalized gluon field $A^{\mu}_r$ and the renormalized coupling constant $g_s$. The renormalized quantities are related to their corresponding bare quantities as
\begin{eqnarray}
&& m_Q^0=Z_m \,m_Q, ~~\Psi_0=\sqrt{Z_2} \,\Psi_r, \nonumber \\
&&A^{\mu}_0=\sqrt{Z_3} \,A^{\mu}_r,~~~g_s^0= Z_g \,g_s \, ,
\end{eqnarray}
where the renormalization constants $Z_i = 1 + \delta Z_i$, with $i= m, 2, 3, g$, respectively. The quantities $\delta Z_i$ are fixed by the renormalized conditions which define a renormalization scheme. The quark field, quark mass and gluon field are renormalized in the on-mass-shell scheme (OS), whereas the strong coupling constant $g_s$ is renormalized in the $\overline{\rm MS}$ scheme. The expressions of $\delta Z_i$ can be derived:
\begin{eqnarray}
\label{rencont}
\delta Z^{OS}_m&=&-3~C_F \frac{\alpha_s}{4\pi}\left[\frac{1}{\epsilon_{UV}}- \gamma_E+
 {\rm ln}\frac{4\pi \mu_R^2}{m^2}+\frac{4}{3}\right],\nonumber\\
\delta Z^{OS}_2&=&-C_F \frac{\alpha_s}{4\pi}\left[\frac{1}{\epsilon_{UV}}+ \frac{2}{\epsilon_{IR}}-3~\gamma_E+3~ {\rm ln}\frac{4\pi \mu_R^2}{m^2}+4\right], \nonumber\\
 \delta Z^{OS}_3&=&\frac{\alpha_s}{4\pi}\left[(\beta'_0-2C_A)\left(\frac{1}{\epsilon_{UV}}-\frac{1}{\epsilon_{IR}}\right) \right. \nonumber\\
 &&\left.-\frac{4}{3}T_F \left(\frac{1}{\epsilon_{UV}}-\gamma_E + {\rm ln}\frac{4\pi \mu_R^2}{m_c^2}\right)\right. \nonumber\\
 &&\left.-\frac{4}{3}T_F \left(\frac{1}{\epsilon_{UV}}-\gamma_E + {\rm ln}\frac{4\pi \mu_R^2}{m_b^2}\right)\right], \nonumber\\
 \delta Z^{\overline{MS}}_g&=&- \frac{\beta_0}{2}\frac{\alpha_s}{4\pi}\left[\frac{1}{\epsilon_{UV}}- \gamma_E+ {\rm ln}~(4\pi) \right],
\end{eqnarray}
where $\mu_R$ is the renormalization scale, $\beta_0=11C_A/3-4 T_F n_f/3$ is the one-loop coefficient of the $\beta$-function in QCD, and $n_f$ is the number of active quark flavors. $\beta'_0=11C_A/3-4 T_F n_{lf}/3$, and $n_{lf} = 3$ is the number of the light-quark flavors. For $SU_{c}(3)$ group, we have $C_A=3$, $C_F=4/3$ and $T_F=1/2$.

The operator products in the definition of the fragmentation functions also require renormalization~\cite{Collins, Mueller}, whose counter-terms in $\overline{\rm MS}$ scheme can be written as~\cite{Braaten2}
\begin{eqnarray}
D^{\rm CT,operator}_{Q\to Q\bar{Q}[^3S_1^{[1]}]}(z)=&&-\frac{\alpha_s}{2\pi}\left[\frac{1}{\epsilon_{UV}}- \gamma_E+ {\rm ln}~(4\pi)+{\rm ln}\frac{\mu_R^2}{\mu_F^2} \right]\nonumber \\
&& \int_z^1 \frac{dy}{y}P_{QQ}(y)D_{Q\to Q\bar{Q}[^3S_1^{[1]}]}^{\rm LO}(z/y),
\end{eqnarray}
where $D_{Q\to Q\bar{Q}[^3S_1^{[1]}]}^{\rm LO}(z)$ denotes the LO fragmentation function in $d$-dimensional space time.

\section{Numerical results}

In doing the numerical calculations, the input parameters are taken as follows:
\begin{eqnarray}
&& m_c=1.5 ~{\rm GeV},\;m_b=4.9 ~{\rm GeV}, \;m_{_Z}=91.1876~{\rm GeV},\nonumber \\
&&\vert R_{J/\Psi}(0) \vert^2=0.810{\rm GeV}^3, \;\vert R_{\Upsilon}(0) \vert^2=6.477{\rm GeV}^3.
\label{inpara}
\end{eqnarray}
where the values of $\vert R_{J/\Psi}(0) \vert^2$ and $\vert R_{\Upsilon}(0) \vert^2$ are taken from the potential-model calculations~\cite{pot}. For the strong coupling constant, we adopt the two-loop formula:
\begin{equation}
\alpha_s(\mu)=\frac{4\pi}{\beta_0~{\rm ln}(\mu^2/\Lambda^2_{\rm QCD})}\left[ 1-\frac{\beta_1~{\rm ln}~{\rm ln}(\mu^2/\Lambda^2_{\rm QCD})}{\beta_0^2~{\rm ln}(\mu^2/\Lambda^2_{\rm QCD})}\right],
\end{equation}
where $\beta_1=\frac{34}{3}C_A^2-4C_F T_F n_f-\frac{20}{3}C_A T_F n_f$ is the two-loop coefficient of the $\beta$-function. According to $\alpha_s(m_{_Z})=0.1185$~\cite{pdg}, we obtain $\Lambda_{\rm QCD}^{n_f=5}=0.233 {\rm GeV}$ and $\Lambda_{\rm QCD}^{n_f=4}=0.337 {\rm GeV}$. Then we have $\alpha_s(2m_c)=0.259$, $\alpha_s(3m_c)=0.223$, $\alpha_s(2m_b)=0.180$, and $\alpha_s(3m_b)=0.164$.

\subsection{The fragmentation functions}

\begin{figure}[htbp]
\includegraphics[width=0.5\textwidth]{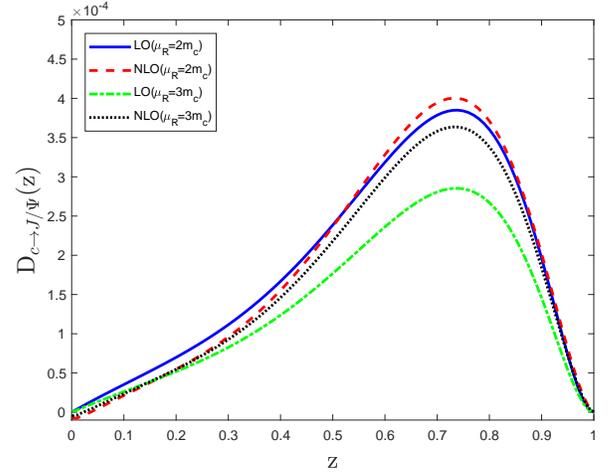}
\caption{The fragmentation function $D_{c\to J/\Psi}(z,\mu_{F},\mu_R)$ as a function of $z$ at an initial factorization scale $\mu_{F}=3m_c$ up to LO and NLO accuracy, respectively. $\mu_R=2m_c$ or $\mu_R=3m_c$. } \label{dzjpsi}
\end{figure}

\begin{figure}[htbp]
\includegraphics[width=0.5\textwidth]{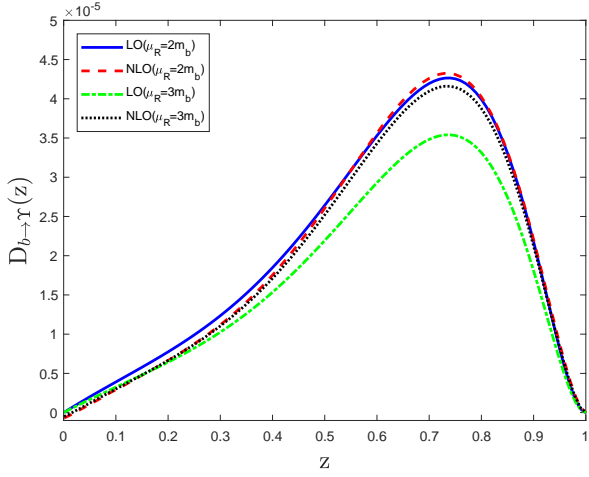}
\caption{The fragmentation function $D_{b\to \Upsilon}(z,\mu_{F},\mu_R)$ as a function of $z$ at an initial factorization scale $\mu_{F}=3m_b$ up to LO and NLO accuracy. $\mu_R=2m_b$ or $\mu_R=3m_b$. } \label{dzupsilon}
\end{figure}

The fragmentation functions for $c\to J/\Psi$ and $b\to \Upsilon$ are shown in Figs.\ref{dzjpsi} and \ref{dzupsilon}, respectively. The initial factorization scale has been set as the minimal invariant mass of the initial $c$ quark or $b$ quark. Two typical values for the renormalization scale $\mu_R$ are adopted, one is the threshold energy to create a $c\bar{c}$ or $b\bar{b}$ pair, and the other is the minimal invariant mass of the initial $c$ quark or $b$ quark. Figs. \ref{dzjpsi} and \ref{dzupsilon} indicate that the renormalization scale dependence for the LO fragmentation functions are large, and the NLO corrections are very important to reduce the renormalization scale dependence.

For future applications, we use polynomials to fit the NLO fragmentation functions. We write the NLO fragmentation functions as the form
\begin{eqnarray}
&&D^{\rm NLO}_{Q\to H}(z,\mu_{F})\nonumber \\
&&= D^{\rm LO}_{Q\to H}(z)\left(1+\frac{\alpha_s(\mu_R)}{2\pi}\beta_0 {\rm ln}\frac{\mu_R^2}{4m_Q^2}\right)\nonumber \\
&&~+\frac{\alpha_s(\mu_R)}{2\pi}{\rm ln}\frac{\mu_{F}^2}{9m_Q^2}\int_z^1 \frac{dy}{y}P_{QQ}(y) D_{Q\to H}^{LO}(z/y)\nonumber \\
&&~+\frac{\alpha_s(\mu_R)^3 \vert R_S(0) \vert^2}{m_Q^3} f(z) ,
\label{eqDzfit}
\end{eqnarray}
where $D^{\rm LO}_{Q\to H}(z)$ is the LO fragmentation function given in Eq.(\ref{lofrag}). For $c\to J/\Psi$, we have
\begin{eqnarray}
f(z)=&&-9.01726z^{10}+18.22777z^9+16.11858z^8\nonumber \\
&&-82.54936z^7+106.57565z^6-72.30107z^5\nonumber \\
&&+28.85798z^4-6.70607z^3+0.84950z^2\nonumber \\
&&-0.05376z-0.00205.
\label{eqfzfit1}
\end{eqnarray}
For $b\to \Upsilon$,
\begin{eqnarray}
f(z)=&&-14.00334z^{10}+46.94869z^9-55.23509z^8\nonumber \\
&&+16.69070z^7+22.09895z^6-26.85003z^5\nonumber \\
&&+13.41858z^4-3.50293z^3+0.46758z^2\nonumber \\
&&-0.03099z-0.00226.
\label{eqfzfit2}
\end{eqnarray}

\begin{table}[htb]
\begin{tabular}{|c |c |c |c |c|}
\hline
    & ~ $P^{\rm LO}\times 10^{4}$~ & ~ $P^{\rm NLO}\times 10^{4}$~ & ~$\langle z \rangle^{\rm LO}$~ & ~ $\langle z \rangle^{\rm NLO}$ ~ \\
\hline
$\mu_R=2 m_c$ &  1.88 & 1.86  & 0.62 &   0.63 \\
$\mu_R=3 m_c$ & 1.40 & 1.72  & 0.62 & 0.63 \\
\hline
\end{tabular}
\caption{The fragmentation probability and the average value of the energy fraction for $c \to J/\Psi$. $\mu_{F}=3m_c$.}
\label{tpsifp}
\end{table}

\begin{table}[htb]
\begin{tabular}{|c | c |c |c |c|}
\hline
 &  ~$P^{\rm LO}\times 10^{5}$~ & ~ $P^{\rm NLO}\times 10^{5}$~ & ~$\langle z \rangle^{\rm LO}$~ & ~ $\langle z \rangle^{\rm NLO}$~ \\
\hline
$\mu_R=2 m_b$ &  2.09 & 2.05  & 0.62 &   0.63 \\
$\mu_R=3 m_b$ & 1.73 & 1.99  & 0.62 & 0.63 \\
\hline
\end{tabular}
\caption{The fragmentation probability and the averaged energy fraction for $b \to \Upsilon$. $\mu_{F}=3m_b$.}
\label{tupfp}
\end{table}

Using the fragmentation function, we can obtain two useful quantities, i.e. the fragmentation probability ($P$) and the averaged energy fraction ($\langle z \rangle$), which are defined as
\begin{eqnarray}
P_{Q\to H} &=& \int_0^1 D_{Q\to H}(z,\mu_{F})\,dz,\\
\langle z \rangle &=& \frac{\int_0^1 z\,D_{Q\to H}(z,\mu_{F})\,dz}{\int_0^1 D_{Q\to H}(z,\mu_{F})\,dz}.
\end{eqnarray}
The fragmentation probabilities and the averaged energy fractions for $c \to J/\Psi$ and $b \to \Upsilon$ are shown in Tables \ref{tpsifp} and \ref{tupfp}, respectively.

\begin{figure}[htbp]
\includegraphics[width=0.5\textwidth]{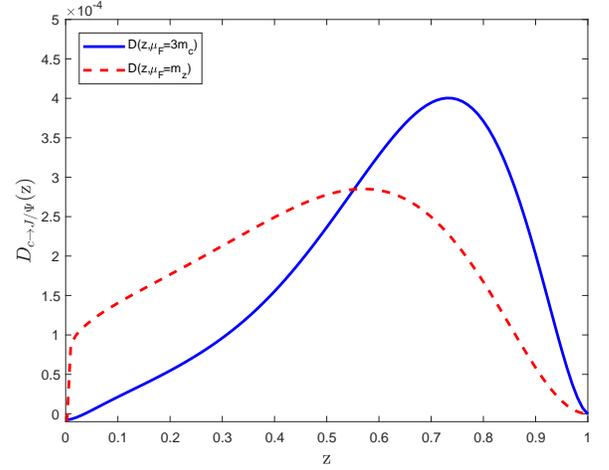}
\caption{The NLO fragmentation function $D_{c\to J/\Psi}$ as a function of $z$ for $\mu_{F}=3m_c$ and $\mu_{F}=m_Z$, respectively. $\mu_R=2m_c$.} \label{dzmzj}
\end{figure}

\begin{figure}[htbp]
\includegraphics[width=0.5\textwidth]{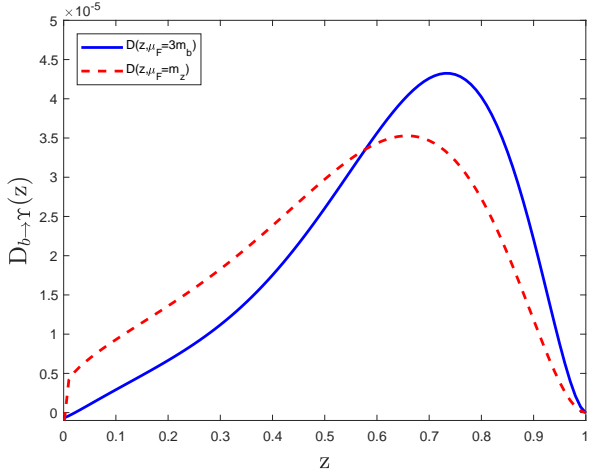}
\caption{The NLO fragmentation function $D_{b\to \Upsilon}$ as a function of $z$ for $\mu_{F}=3m_b$ and $\mu_{F}=m_Z$, respectively. $\mu_R=2m_b$.} \label{dzmzu}
\end{figure}

The fragmentation functions shown in Figs. \ref{dzjpsi} and \ref{dzupsilon} are for $\mu_F=3m_Q$. The fragmentation functions at any other factorization scales can be obtained through DGLAP equation. To apply the NLO fragmentation functions to the production of heavy quarkonia at a $Z$ factory, we present the fragmentation functions for $\mu_F=m_{_Z}$ in Figs. \ref{dzmzj} and \ref{dzmzu}, which are obtained by using the fragmentation functions at the initial value $\mu_{F}=3m_Q$. For definiteness, we set the renormalization scale $\mu_R=2m_{Q}$. In doing the numerical calculation, the DGLAP equation is solved through the Mellin transformation~\cite{evolution1, evolution2}, and the NLO expression for $P_{QQ}$ is used as the evolution kernel. Figs. \ref{dzmzj} and \ref{dzmzu} show that the peaks of the fragmentation functions become lower and shift to a smaller $z$ value for a larger factorization scale, leading to a smaller value for the averaged energy fraction $\langle z \rangle$. For examples, we obtain $\langle z \rangle^{\rm NLO+NLL}|_{\mu_{F}=m_Z}=0.47$ for $D_{c\to J/\Psi}$ and $\langle z \rangle^{\rm NLO+NLL}|_{\mu_{F}=m_Z}=0.54$ for $D_{b\to \Upsilon}$.

\subsection{The $J/\Psi$ and $\Upsilon$ production at a $Z$ factory}
\label{sec4}

As an application, we apply the NLO fragmentation functions of $c \to J/\Psi$ and $b \to \Upsilon$ to the production of $J/\Psi$ and $\Upsilon$ at a super $Z$ factory.

A Chinese group has raised the proposal of constructing a high-luminosity $e^+ e^-$ collider in China, the so-called super $Z$ factory~\cite{zfactory}, which is similar to the Gigaz program suggested by the Internal Linear Collider Collaboration~\cite{gigaz1, gigaz2} but with a even higher luminosity. As for this purposed super $Z$ factory, an $e^+ e^-$ collider shall run at the energies around the mass of $Z^0$-boson resonance and with a high luminosity up to ${\cal L}=10^{34}-10^{36} {\rm cm}^{-2} {\rm s}^{-1}$. Due to the $Z^0$-boson resonance effect, large numbers of $J/\Psi$ and $\Upsilon$ events can be produced, thus providing a good platform studying the $J/\Psi$ and $\Upsilon$ properties.

In this case, the factorization formula for the production of $J/\Psi$ or $\Upsilon$ can be written as
\begin{eqnarray}
\frac{d\sigma_{e^+e^-\to H+X}}{dz}=&&\sum_i \int_z^1 \frac{dy}{y}\frac{d\hat{\sigma}_{e^+e^-\to i+X}(y,\mu_F)}{dy}\cdot \nonumber \\
&& \quad\quad D_{i\to H}(z/y,\mu_F),\label{eqfact}
\end{eqnarray}
where the energy fraction $z$ is defined as $z\equiv 2 p \cdot k/k^2$, $p$ is the momentum of the produced quarkonium, $k$ is the sum of the momenta of the initial electron and positron. Up to NLO level, the parton $i$ may be a heavy quark or a heavy antiquark. For the quarkonium production, the fragmentation function $D_{\bar{Q}\to H}$ is the same as $D_{Q\to H}$.

Due to that the coefficient functions $d\hat{\sigma}/dy$ are independent to the species of the produced hadron, they can be extracted by applying the pQCD factorization formula to the production of an on-shell heavy quark ($Q$) or heavy antiquark ($\bar{Q}$)~\cite{pqcdfac1, pqcdfac2}. The expression for the coefficient function $d\hat{\sigma}/dy$ in $\overline{\rm MS}$-scheme up to NLO level has been given in Refs.\cite{coefun1, coefun2}, e.g.
\begin{eqnarray}
&&\frac{d\hat{\sigma}_{e^+e^-\to Q+X}^{\rm NLO}}{dy}(y,\mu_F)\nonumber \\
&&=\sigma_{e^+e^- \to Q \bar{Q}}^{\rm LO}\left[\delta(1-y)+\frac{\alpha_s(\mu_R)}{2\pi} \left( P_{QQ}(y) {\rm ln}\frac{s}{\mu_F^2}+C(y)\right) \right],\nonumber \\
\label{eqcoefunc}
\end{eqnarray}
where $\sigma_{e^+e^- \to Q \bar{Q}}^{\rm LO}$ is the LO cross section for the $Q\bar{Q}$ production. And in the massless limit $m_Q \to 0$~\footnote{Because the coefficient function $d\hat{\sigma}/dy$ is infrared safe, so we can take the limit $m_Q \to 0$ to do our calculation, which shall introduce a small error of ${\cal O}(m_Q^2/s)$.}, we have
\begin{eqnarray}
\sigma_{e^+e^- \to Q \bar{Q}}^{\rm LO}=&& \frac{4\pi N_c \alpha^2}{3s}\left[e_e^2 e_Q^2+2e_e v_e e_Q v_Q \rho_1(s)\right. \nonumber \\
&&  \left. +(v_e^2+a_e^2)(v_Q^2+a_Q^2) \rho_2(s)\right],
\end{eqnarray}
where $e_f$ is the electric charge of fermion $f$,
\begin{eqnarray}
v_f&=&(T_{3f}-2e_f {\rm sin}^2\theta_w)/(2{\rm sin}\theta_w {\rm cos}\theta_w),  \\
a_f&=&T_{3f}/(2{\rm sin}\theta_w {\rm cos}\theta_w),
\end{eqnarray}
are the vector and axial-vector couplings of fermion $f$ to the $Z$ boson, $T_{3f}$ is the third component of weak isospin of the fermion $f$, $\theta_w$ is the weak mixing angle, and the propagator functions are:
\begin{eqnarray}
\rho_1(s)&=&\frac{s(s-m_z^2)}{(s-m_Z^2)^2+m_Z^2 \Gamma_Z^2},\nonumber \\
\rho_2(s)&=&\frac{s^2}{(s-m_Z^2)^2+m_Z^2 \Gamma_Z^2}.
\end{eqnarray}
The function $C(y)$ in the massless limit $m_Q \to 0$ takes the form
\begin{eqnarray}
C(y)=&&C_F\left\{\left(\frac{2\pi^2}{3}-\frac{9}{2}\right)\delta(1-y)-\frac{3}{2}\left(\frac{1}{1-y}\right)_+ \right. \nonumber \\
&&  +2\left(\frac{{\rm ln}(1-y)}{1-y} \right)_+ -(1+y)\left[2 {\rm ln}y+  {\rm ln}(1-y) \right]\nonumber \\
&&\left.+4\frac{{\rm ln} \,y}{1-y} +\frac{5}{2}-\frac{3y}{2} \right\}.
\end{eqnarray}

For comparison, we adopt three strategies to calculate the differential cross sections $d\sigma/dz$ under the fragmentation approach. We denote them as ``Frag, LO", ``Frag, NLO" and ``Frag, NLO+NLL", respectively. For the case of ``Frag, LO", the differential cross sections are calculated through
\begin{eqnarray}
&& \frac{d\sigma^{\rm LO}_{e^+e^-\to H+X}}{dz}\nonumber\\
&=& 2\int_z^1 \frac{dy}{y}\frac{d\hat{\sigma}^{\rm LO}_{e^+e^-\to Q+\bar{Q}}}{dy}(y,\mu_{F})\cdot D^{\rm LO}_{Q\to H}(z/y,\mu_F) \nonumber \\
&=& 2 \sigma^{\rm LO}_{e^+e^-\to Q+\bar{Q}}\cdot D^{\rm LO}_{Q\to H}(z,\mu_F)
\label{eqfactLO}
\end{eqnarray}
where the factor $2$ comes from that the contribution of the $\bar{Q}$ fragmentation which is the same as that of $Q$ fragmentation. $D^{LO}_{Q\to H}(z)$ denotes the LO fragmentation function which has been given in Eq.(\ref{lofrag}). In the second equation, we have used the fact that $d\hat{\sigma}^{\rm LO}_{e^+e^-\to Q+\bar{Q}}/dy=\sigma^{\rm LO}_{e^+e^-\to Q+\bar{Q}} \delta(1-y)$. For the case of ``Frag, NLO", the differential cross sections are calculated through
\begin{eqnarray}
\frac{d\sigma^{\rm NLO}_{e^+e^-\to H+X}}{dz}=&& 2\int_z^1 \frac{dy}{y}\frac{d\hat{\sigma}^{\rm NLO}_{e^+e^-\to Q+X}}{dy}(y,\mu_{F}) \nonumber \\
&&\times D^{\rm NLO}_{Q\to H}(z/y,\mu_{F}).\label{eqfactNLO}
\end{eqnarray}
In the above calculation, the factorization and renormalization scales are set as $\mu_{F}=3m_Q$ and $\mu_R=2m_Q$. For the case of ``Frag, NLO+NLL", the differential cross sections are calculated through
\begin{eqnarray}
\frac{d\sigma^{\rm NLO+NLL}_{e^+e^-\to H+X}}{dz}=&& 2\int_z^1 \frac{dy}{y}\frac{d\hat{\sigma}^{\rm NLO}_{e^+e^-\to Q+X}}{dy}(y,\mu_{F}) \nonumber \\
&&\times D_{Q\to H}(z/y,\mu_{F}), \label{eqfactNLOLL}
\end{eqnarray}
where the factorization scale and renormalization scale are set as $\mu_F=\mu_R=m_{_Z}$, and the fragmentation functions $D_{Q\to H}(z,\mu_{F}=m_{_Z})$ are obtained through solving the DGLAP evolution equation, i.e., Eq.(\ref{dglap}) and the initial fragmentation functions $D^{\rm NLO}_{Q\to H}(z,\mu_{F}=3m_Q)$ with $\mu_R=2m_Q$ are used as the boundary condition. This way, the large log-terms such as $\ln(m_Q^2/m_Z^2)$ are resummed up to next-to-leading-log (NLL) accuracy.

\begin{figure}[htb]
\includegraphics[width=0.5\textwidth]{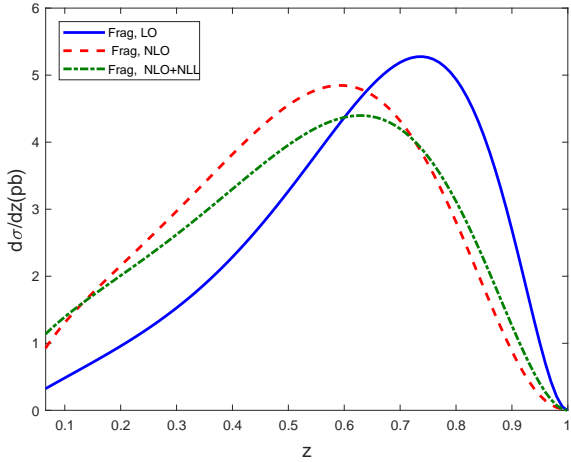}
\caption{The differential cross section $d\sigma/dz$ of $e^+e^- \to J/\Psi+X$ at the $Z$ pole under the fragmentation approach.} \label{sigmazc1}
\end{figure}

\begin{figure}[htb]
\includegraphics[width=0.5\textwidth]{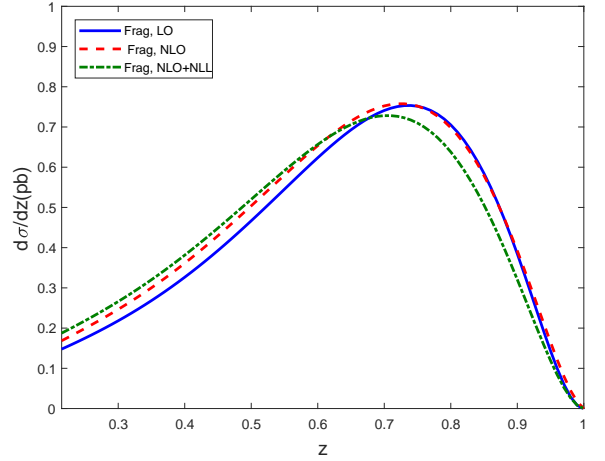}
\caption{The differential cross section $d\sigma/dz$ of $e^+e^- \to \Upsilon+X$ at the $Z$ pole under the fragmentation approach.} \label{sigmazb1}
\end{figure}

We present the differential cross sections $d\sigma/dz$ for the production of $J/\Psi$ and $\Upsilon$ via the fragmentation approach in Figs. \ref{sigmazc1} and \ref{sigmazb1}. In drawing the figures, we have omitted the $\gamma-\gamma$ and $\gamma-Z$ contributions, which are quite small compared with the dominant $Z-Z$ contribution around the $Z$ pole. Figs. \ref{sigmazc1} and \ref{sigmazb1} show how the NLO-terms and the leading and next-to-leading logarithms affect the predictions. For the $J/\Psi$ production, the NLO contribution is significant, and after including the NLO-terms, the distribution becomes softer and the value of $z$ corresponding to the peak value of the distribution becomes smaller. For the $\Upsilon$ production, the NLO contribution is relatively small compared with the $J/\Psi$ case.

\begin{table}[htb]
\begin{tabular}{|c |c |c |c|}
\hline
~~ &  ~Frag, LO~ & ~Frag, NLO~ & ~Frag, NLO+NLL \\
\hline
$J/\Psi$ &  2.58 & 2.77  & 2.65  \\
$\Upsilon$ & 0.368 & 0.382  & 0.377  \\
\hline
\end{tabular}
\caption{Total cross sections (in unit: $pb$) of production channels $e^+e^- \to J/\Psi+X$ and $e^+e^- \to \Upsilon+X$ at the $Z$ pole.}
\label{tcs}
\end{table}

Integrating the differential cross sections $d\sigma/dz$ over $z$, we can obtain the total cross sections for $J/\Psi$ and $\Upsilon$ production at the $Z$ factory, which can be simplified as
\begin{eqnarray}
\sigma(H) &=& 2 \, P_{Q\to H}\int_0^1 \frac{d\hat{\sigma}_{e^+e^- \to Q+X}}{dy}(y,\mu_F),
\end{eqnarray}
where $P_{Q\to H}$ is the fragmentation probability for $Q$ into the quarkonium $H$. The results are presented in Table \ref{tcs}, which shows that the NLO corrections enhance the total cross section $\sim 7\%$ for $J/\Psi$ and $\sim 4\%$ for $\Upsilon$.

\subsection{A comparison of $J/\Psi$ and $\Upsilon$ production via the $Z$ decays up to NLO level}

\begin{figure}[htbp]
\includegraphics[width=0.5\textwidth]{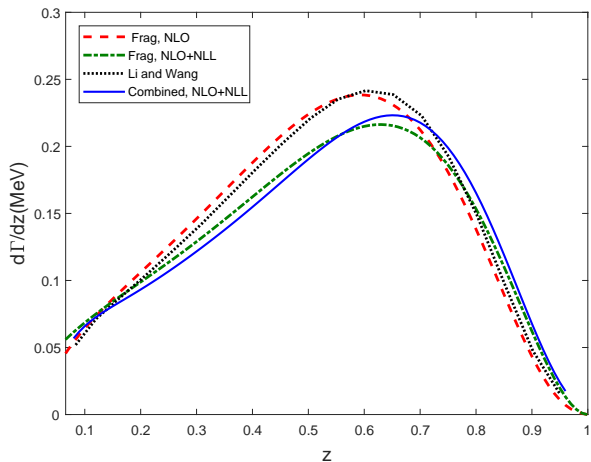}
\caption{The differential width $d\Gamma/dz$ of $Z \to J/\Psi+X$.} \label{gammazc1}
\end{figure}

\begin{figure}[htbp]
\includegraphics[width=0.5\textwidth]{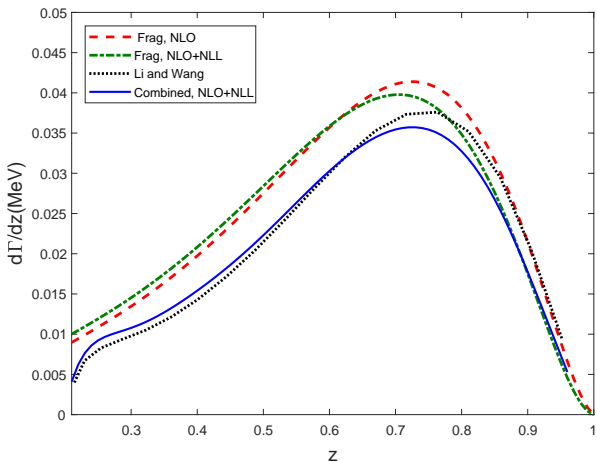}
\caption{The differential width $d\Gamma/dz$ of $Z \to \Upsilon+X$.} \label{gammazb1}
\end{figure}

As a final remark, we compare the NLO results under the fragmentation function approach and the complete pQCD approach. The complete NLO pQCD calculations of $Z \to J/\Psi+X$ and $Z \to \Upsilon+X$ have been done by Ref.\cite{JXWang}. We present the differential decay widths $d\Gamma/dz$ for $Z \to J/\Psi+X$ and $Z \to \Upsilon+X$ in Figs. \ref{gammazc1} and \ref{gammazb1}, in which all the input parameters are taken as those of Ref.\cite{JXWang} and the results denoted by ``Li and Wang" are results under the complete NLO calculation. The results denoted by ``Combined, NLO+NLL" are results combining the results from the complete pQCD approach which are taken from Ref.\cite{JXWang} and the results from the fragmentation approach, i.e, $d\Gamma/dz|_{\rm Combined, NLO+NLL}=d\Gamma/dz|_{\rm Li\; and\; Wang}+(d\Gamma/dz|_{\rm Frag, NLO+NLL}-d\Gamma/dz|_{\rm Frag, NLO})$. For the case of $J/\Psi$ production at the $Z$ factory, Fig.\ref{gammazc1} shows the fragmentation contributions dominant the decay width, since the ``Frag, NLO" shape is very close to the complete NLO one. Moreover, the total decay widths are $\Gamma(Z\to J/\Psi+X)|_{\rm Frag, NLO}=136\,{\rm keV}$, $\Gamma(Z\to J/\Psi+X)|_{\rm Frag, NLO+NLL}=130 \,{\rm keV}$, $\Gamma(Z\to J/\Psi+X)|_{\rm Li\; and\; Wang}=136\,{\rm keV}$ and $\Gamma(Z\to J/\Psi+X)|_{\rm Combined,NLO+NLL}=130\,{\rm keV}$. Fig.\ref{gammazb1} shows that the ``Frag, NLO" result is larger than the result under the complete NLO calculation at small $z$-region, thus the fragmentation approximation for the $\Upsilon$ production is not as good as the $J/\Psi$ case. This is because the $b$-quark mass is larger than $c$-quark mass, and the power correction in $m_b^2/s$ for the case of $\Upsilon$ is larger than the case of $J/\Psi$. However, the combined result counts both the large power correction and the large logarithms, then give a good prediction. Moreover, the total decay widths are $\Gamma(Z\to \Upsilon+X)|_{\rm Frag, NLO}=20.9 \,{\rm keV}$, $\Gamma(Z\to J/\Upsilon+X)|_{\rm Frag, NLO+NLL}=20.6\,{\rm keV}$, $\Gamma(Z\to J/\Upsilon+X)|_{\rm Li\; and\; Wang}=17.38\,{\rm keV}$, and $\Gamma(Z\to J/\Upsilon+X)|_{\rm Combined,NLO+NLL}= 17.08\,{\rm keV}$.

\section{Summary}

In the present paper, we have calculated the fragmentation function for a heavy quark into heavy quarkonium, e.g. $c\to J/\Psi$ or $b\to \Upsilon$, up to NLO level. Our present results are complementary to the previous works on the fragmentation function of a gluon into heavy quarkonia done in the literature, which is pQCD calculable due to the fact that the gluon should be hard enough to form a heavy quark-and-antiquark pair.

Our results show that the NLO correction is important to suppress the renormalization scale uncertainty and to achieve a reliable fragmentation prediction. Our calculations are based on the gauge-invariant definition of the fragmentation function suggested by Collins and Soper. To avoid large logarithms appearing in the perturbative series of the fragmentation function, we first derive the fragmentation function at an initial (reasonable) factorization scale $\mu_{F}=3m_Q$, and then run to any factorization scale with the help of the DGLAP evolution equation. This treatment, in effect, resums the large logarithms and forms a reliable prediction. Thus for the cases when the fragmentation dominants the quarkonium productions or decays, our present calculated fragmentation functions shall be of great help for a more precise pQCD prediction. As an application, we have applied the obtained fragmentation functions to the production of $J/\Psi$ and $\Upsilon$ at the super $Z$ factory. The shape of the $J/\Psi$ distribution changes significantly by introducing the NLO corrections, and the total cross section increases by $\sim 7\%$. The shape of the $\Upsilon$ distribution changes slightly by introducing the NLO corrections, and the total cross section increases only $\sim 4\%$.

\noindent {\bf Acknowledgments:} This work was supported in part by the Natural Science Foundation of China under Grant No.11625520, No.11847222, No.11847301, No.11675239, No.11535002, and by the Fundamental Research Funds for the Central Universities under Grant No.2019CDJDWL0005.

\end{document}